\documentclass[a4paper,11pt]{article}

\pdfoutput=1  

\usepackage{graphics,psfrag}
\usepackage{amsmath,amsthm,amssymb,epsfig,euscript,array,cite,cancel,color}
\usepackage{mathrsfs}
\usepackage{cases,empheq}
\usepackage{bm}
\usepackage[colorlinks=false]{hyperref}
\usepackage{verbatim}
\usepackage{ulem}
\usepackage[vcentermath]{youngtab}   
\usepackage[usenames,dvipsnames]{xcolor}  
\usepackage[font=small]{caption} 

\usepackage[paper=a4paper,dvips,top=2.54cm,left=2.74cm,right=2.34cm,
    foot=1cm,bottom=2.54cm]{geometry}

\newcommand{\ba}{\begin{eqnarray}}
\newcommand{\ea}{\end{eqnarray}}
\newcommand{\nn}{\nonumber}

\def \be {\begin{equation}}
\def \ee {\end{equation}}
\def \barr {\begin{array}}
\def \earr {\end{array}}
\def \bea {\begin{eqnarray}}
\def \eea {\end{eqnarray}}

\def \ble {\begin{widetext}\begin{equation}}
\def \ele {\end{equation}\end{widetext}}
\def \blea {\begin{widetext}\begin{eqnarray}}
\def \elea {\end{eqnarray}\end{widetext}}

\def \nn {\nonumber}

\newcommand{\eq}[1]{(\ref{#1})}

\def \d {\delta}

\def \m {\mu}
\def \n {\nu}

\def \l {\lambda}
\def \L {\Lambda}

\def \and {{\textrm{and}}}

\author{Kilar Zhang\footnote{kilar.zhang@gmail.com} \;and 
Feng-Li Lin\footnote{linfl@gapps.ntnu.edu.tw $\quad$ Corresponding Author}
\\
\\
\small\it Department of Physics, National Taiwan Normal University, Taipei 11677, Taiwan
}

\title{\bf {\Large Constraint on hybrid stars with gravitational wave events} }

\date{}
\begin{document}

\maketitle

\thispagestyle{empty}
 
\abstract{

\hspace{2mm}
Motivated by the recent discoveries of compact objects from LIGO/Virgo observations, we~study the possibility of identifying some of these objects as compact stars made of dark matter called dark stars, or the mix of dark and nuclear matters called hybrid stars. In particular, in~GW190814, a new compact object with 2.6 $M_{\odot}$ is reported. This could be the lightest black hole, the heaviest neutron star, and a dark or hybrid star.  In this work, we extend the discussion on the interpretations of the recent LIGO/Virgo events as hybrid stars made of various self-interacting dark matter (SIDM) in the isotropic limit. We pay particular attention to the saddle instability of the hybrid stars which will constrain the possible SIDM models.
 }

\thispagestyle{empty}
\newpage

\tableofcontents

\section{Introduction}

The LIGO/Virgo events GW170817~\cite{TheLIGOScientific:2017qsa,Abbott:2018wiz} and GW190425~\cite{Abbott:2020uma} are in general considered as binaries of neutron stars (BNSs), as~well as the possibility of being binaries of hybrid stars. However, in~\cite{Abbott:2020khf}, a~new binary coalescence event GW190814 is reported, while one is a black hole with 23 $M_{\odot}$, and~the other is a compact object with only 2.6 $M_{\odot}$ (with a range between 2.50 and 2.67$M_{\odot}$). Then, the~question is what is the identity of the secondary compact object? There are several possibilities: It could be the lightest black hole, or~the heaviest neutron star, or~a dark star. In~addition, there are also possibilities for this compact object to be a hybrid star made of neutron and dark matter~\cite{Zhang:2020pfh}.  Below, we will extend a few more discussions on the above possibilities, and~then focus on elaborating more on the scenarios of dark and hybrid stars in the rest of the~paper.
 
{ {The scenario of black hole---}
%
}The black holes observed by LIGO/Virgo are all known to have masses more than 5 $M_{\odot}$. However, the~remnant of GW170817, which is identified to be the BNS by parameter estimation (PE), is estimated to have a mass of about 2.7 $M_{\odot}$, and~is believed to be a black hole~\cite{Abbott:2018wiz}. Thus, the~black hole of mass less than 5 $M_{\odot}$ can be formed via the coalescence of BNS. 
Since no information about tidal Love number is given in the released PE of GW190814, and~no electromagnetic follow-up is reported, we cannot exclude the possibility that this 2.6 $M_{\odot}$ object is the lightest black hole, which could likely be either formed astronomically via the coalescence of BNS, or~formed in the early universe as a primordial black hole~\cite{Yuan:2020iwf}. Alternatively, the~light black hole may come from core-collapse supernova of a massive normal star~\cite{Belczynski:2011bn}.

{The scenario of neutron star---}Before this GW190814 event, masses of neutron stars are generally considered to be less than 2.3 $M_{\odot}$ \cite{Margalit:2017dij, Rezzolla:2017aly}. In~fact, within~our galaxy, the~most massive known pulsar has $2.14^{+0.10}_{-0:09} M_{\odot}$ at $68.3\%$ credible interval measured by the Shapiro time-delay of its white dwarf companion~\cite{Cromartie:2019kug}. Most known equations of state (EoSs) for the dense nuclear matter used in the astronomical search of neutron stars cannot sustain such high mass as 2.3 $M_{\odot}$, although~this can also be seen as a result of tuning the parameters of the EoSs to not violate the upper mass limit of the astronomical observations.
On the other hand, rapid spinning of neutron star~\cite{Breu:2016ufb,Most:2020bba} can increase the maximal mass up to $\sim$20\%, which will lead to $\sim$2.7 $M_{\odot}$. This case cannot be totally ruled out since there is no constraint on the spin of the secondary object in GW190814. However, such high spin NSBH system may hardly exist because of the expected subsequent collapse into black hole due to the loss of spin via gravitational wave~radiation.

{The scenario of dark star and hybrid star---}Currently, dark stars have not yet been confirmed in observations so that it remains to pin down the accessible model space of dark matter and the associated EoSs by the future observations in the coming era of gravitational wave astronomy. In~light of the varieties of dark matter models, it is easy for the resultant dark stars to cover a wide mass range, say 1 to 5 $M_{\odot}$, or~even one order higher with proper parameters. This makes the dark stars or the hybrid stars of dark and nuclear matters to be highly possible candidates to explain the companion star of GW190814. In~the remaining of this paper, we will explore this possibility by studying the mass--radius relations for the dark and hybrid stars based on various dark matter models inspired by the particle physics. Specifically, we consider the massive bosonic field $\phi$ with the following self-interactions, $\phi^4$, $\phi^6$, their linear combinations and $\phi^{10}$. 

One way to characterize a compact star is through its Tidal Love Number (TLN)\footnote{The TLN denoted by $\Lambda$ is defined by
$
Q_{ab}=- M^5  \Lambda \; {\cal E}_{ab},
$ 
 where $M$ is the mass of the star, $Q_{ab}$ is the induced quadrupole moment, and~${\cal E}_{ab}$ is the external gravitational tidal field strength.} \cite{Hinderer:2007mb,Postnikov:2010yn}, which~shows the star’s tendency to deform under an external quadrupolar tidal field. 
It is known that the TLN of black holes in Einstein gravity is vanishing, and~the overall TLN effect for a compact binary coalescence event is a weighted average of the individual TLNs. Thus, for~GW190814, the~overall tidal effect is insignificant in the resultant gravitational waveform due to the high mass ratio between black hole and the companion star. For~example, even if the 2.6 $M_{\odot}$ object has a large TLN such as 30,000, when combining with zero TLN of the companion black hole of 23 $M_{\odot}$, the~overall TLN $\tilde \Lambda$\footnote{$\tilde{\Lambda}={16\over 13}{(M_1+12 M_2) M_1^4 \Lambda_1+(M_2+12 M_1) M_2^4 \Lambda_2 \over (M_1+M_2)^5}$ where $M_i$ and $\Lambda_i$ with $i=1$ or $2$ is the mass and TLN for the $i$-th component compact object.} contributed to the gravitational waveform is about 43, which can hardly be observed. Since there is no available information on TLN from this event anyway, we need more future events with much smaller masses where the TLN judgement can be applied. For~the above reason, we will not present the mass-TLN relation in this~work. 

The rest of the paper is organized as follows. In~the next section, we will sketch how we extract the EoSs of the bosonic dark matter models in the isotropic limit. In~Section~\ref{sec 3}, we discuss the stability of the dark and hybrid stars based on the famous Bardeen--Thorne--Meltzer (BTM) criteria, especially with the emphasis on the saddle instability when the mixed phase rule does not apply. The~key result is exposed in Section~\ref{sec 4} by plotting the mass--radius relation for various EoSs extracted from the respective dark matter models. Based on the mass--radius relation, we discuss the relevance to and interpretation of GW190814. Finally, we conclude our paper in Section~\ref{sec 5}. 
\section{EoS for Bosonic SIDM in the Isotropic~Limit} \label{sec 2}

Most of the dark matter models are motivated by particle physics, which have either weak or no interaction with the standard model particles. The~former is called the WIMP, namely weak interacting massive particles, and~the latter will also include the self-interactions to explain the core profile of dark halos well so that it is usually called SIDM, self-interacting dark matter.  In~this paper, we will mostly focus on SIDM. The~simplest model of SIDM is the massive $\phi^4$ bosonic field theory considered in~\cite{Colpi:1986ye}. Naively, one should solve the combined scalar-tensor field equations for the possible compact dark star configurations. However, if~we assume the scalar profile inside the star varies very slowly, we can neglect the spatial profile and obtain an isotropic dark star configurations. In~the low energy regime, these kinds of isotropic configurations are more favored energetically than the nonisotropic ones. Thus, for~simplicity, we will only consider such kind of dark and hybrid stars. One additional advantage for solving the isotropic dark stars is we can first extract the EoS in the isotropic limit, then solve the Tolman--Oppemheimer--Volkoff (TOV) equations for the mass--radius relation. This is numerically far easier than solving the scalar-tensor field equations by the shooting method\footnote{A way for solving boundary value problems by changing them into initial value problems. One 'shoot' out trajectories in different directions until a trajectory with the desired boundary value is found, which usually costs a long machine time.}. 

In the following, we sketch how to extract the EoS for the generic bosonic SIDM models by generalizing the procedure given in~\cite{Colpi:1986ye}. The~Lagrangian of the bosonic SIDM model considered in this work is  
\be\label{effective}
{\cal L}=-{1\over 2} g^{\mu\nu} \partial_{\mu}\phi^* \partial_{\nu} \phi -V(|\phi|)
\ee
from which we can obtain the field equation
\ba\label{scalar-eq}
0=\partial_\mu (\sqrt{-g}\partial^\mu \phi )-\sqrt{-g} V'(\phi),
\ea
where $V':=\partial_{\phi}V$.

We will consider the following metric ansatz with spherical symmetry for the dark or hybrid star
\ba\label{metrica}
ds^2=-B(r) d t^2+A(r) d r^2 + r^2 d \Omega.
\ea

This metric is sourced by the spherically symmetric scalar field configuration
\ba\label{fullscalar}
\phi(r,t)=\Phi(r)e^{-i \omega t},
\ea
which should solve \eq{scalar-eq} in the space-time \eq{metrica}, i.e.,~
\ba
0&=&\partial_r (r^2 \sqrt{AB}\frac{1}{A}\partial_r\Phi )+\omega^2
(r^2 \sqrt{AB}\frac{1}{B}\Phi )-r^2 \sqrt{AB} V'(\Phi)\nn\\ 
&=&\partial_r (r^2 \sqrt{\frac{B}{A}}\partial_r\Phi )+r^2 \sqrt{AB}\left[
\frac{\omega^2}{B}\Phi - V'(\Phi)\right].\label{scalar}
\ea

On the other hand, the~stress tensor associated with Lagrangian \eq{effective} is 
\ba
 T_{\n}^{\m}={1\over 2} g^{\mu\sigma} \left(\nabla_{\sigma}\phi^* \nabla_{\nu} \phi
 +\nabla_{\sigma}\phi \nabla_{\nu} \phi^* \right)- \d_{\n}^{\m}({1\over 2} g^{\rho\lambda} \nabla_{\rho}\phi^* \nabla_{\lambda} \phi+V(|\phi|))
\ea
which satisfies the conservation law 
\ba
 \nabla_{\m} T_{\n}^{\m}=0, 
\ea
and also sources the Einstein equation
\be\label{Einstein}
G_{\mu\nu}=8\pi G_N T_{\mu\nu} 
\ee
with $G_{\mu\nu}$ the Einstein tensor and $G_N$ the Newton constant. The~total configurations for a boson star specified by $A(r)$, $B(r)$, and~$\Phi(r)$ should be determined by solving \eq{scalar-eq} and \eq{Einstein} together. In~general, the~stress tensor for the stationary configurations of \eq{fullscalar} satisfying \eq{scalar} in the space-time \eq{metrica} takes the following form in the co-moving frame
\ba\label{Tmunu-1}
T_{\mu}^{\nu}=\textrm{diag}(-\rho,p,p_{\perp},p_{\perp}).
\ea

However, in~the isotropic limit, it will further reduce to the form of a perfect fluid, i.e.,~$p_{\perp}=p$.
 
Now, we consider the following concrete example of bosonic dark matter model with the self-interactions as 
\be\label{Vn}
V(\phi)=\frac12 m^2|\phi|^2+{1\over n} {\lambda_n \over \Phi_0^{n-4}} |\phi|^n\;.
\ee

This can be thought as a model possessing of a UV $Z_n$ symmetry, which is however mildly broken at low energy by the mass term, that is, the~$Z_n$ symmetry is approximately good at low energy if $\Phi_0 \gg m$. For~this bosonic SIDM model, its stress tensor is given by 
\ba
  T_{\n}^{\m}= \left[
 \begin{matrix}
  - {1\over 2}({\omega^2 \over B}+m^2)\Phi^2 - {1\over n} {\lambda_n \over \Phi_0^{n-4}} \Phi^n - \frac{(\partial_{r}\Phi)^2}{2A }  \qquad 0 \qquad 0 \qquad~0\\
   0 \qquad  {1\over 2}({\omega^2 \over B}-m^2)\Phi^2 - {1\over n} {\lambda_n \over \Phi_0^{n-4}} \Phi^n + \frac{(\partial_{r}\Phi)^2}{2A } \qquad 0 \qquad~0\\
   0 \qquad 0\qquad  {1\over 2}({\omega^2 \over B}-m^2)\Phi^2 - {1\over n} {\lambda_n \over \Phi_0^{n-4}} \Phi^n - \frac{(\partial_{r}\Phi)^2}{2A } \qquad~0\\
   0 \qquad 0\qquad 0 \qquad {1\over 2}({\omega^2 \over B}-m^2)\Phi^2 - {1\over n} {\lambda_n \over \Phi_0^{n-4}} \Phi^n - \frac{(\partial_{r}\Phi)^2}{2A }\;
  \end{matrix}
  \right] \label{Tmunu-2}
\ea
which takes the form of \eq{Tmunu-1} as~expected. 

To perform the isotropic limit, we first introduce the following dimensionless variables
\be
x=rm, \qquad \Omega=\frac{\omega}{m}, \qquad \sigma=\sqrt{4\pi }\frac{\Phi}{M_{Pl}}\;.
\ee

Here, $M_{Pl}$ is the Planckian mass scale which is related to the Newton constant by $G_N=1/M^2_{PL}$. Then, the~scalar field equation \eq{scalar} is reduced into 
\ba\label{key}
0
=\partial_x (x^2 \sqrt{\frac{B}{A}}\partial_x\sigma) +x^2 \sqrt{AB}\left[
\frac{\Omega^2}{B}\sigma  - \frac{\sqrt{4\pi }}{ m^2M_{Pl}}V'\left(\frac{M_{Pl}\sigma}{\sqrt{4\pi }}\right)\right].
\ea

By further introducing a new dimensionless parameter $\L_n$ defined by
\be\label{Lambda_n}
\L_n=\Big(\l_n\frac{\Phi_0^2}{m^2}\Big)^{2\over n-2}  \Big(\frac{M_{Pl}}{ \sqrt{4\pi }\Phi_0}\Big)^2
\ee
and the new scaled quantities
\be 
\sigma_*=\sqrt{\L_n}\sigma, \qquad x_*=x/\sqrt{\L_n},
\ee 
then the energy density and pressures of of \eq{Tmunu-2} in the form of \eq{Tmunu-1} can be expressed as follows:
\bea\label{rho-1}
 \rho&=& {m^2 M_{Pl}^2  \over 4\pi\L_n}\Big[ {1\over 2}({\Omega^2 \over B}+1)\sigma_*^2 + {1\over n}\sigma_*^n +\frac1{ \L_n} \frac{(\partial_{x_*}\sigma_*)^2}{2A } \Big], \\
 p&=& {m^2 M_{Pl}^2 \over 4\pi\L_n}\Big[ {1\over 2}({\Omega^2 \over B}-1)\sigma_*^2 - {1\over n}\sigma_*^n +\frac1{ \L_n} \frac{(\partial_{x_*}\sigma_*)^2}{2A }\Big], 
 \\
 p_{\perp}&=& {m^2 M_{Pl}^2 \over 4\pi\L_n}\Big[ {1\over 2}({\Omega^2 \over B}-1)\sigma_*^2 - {1\over n}\sigma_*^n -\frac1{ \L_n} \frac{(\partial_{x_*}\sigma_*)^2}{2A }\Big], \label{pperp-1}
\eea
and \eqref{key} is also further reduced to
\ba\label{reduced key}
0
=\frac1{\sqrt{\L_n}}\partial_{x_*} (x_*^2 \sqrt{\frac{B}{A}}\partial_{x_*}\sigma_*)+\sqrt{\L_n} x_*^2 \sqrt{AB}\left[
(\frac{\Omega^2}{B}-1)\sigma_* - \sigma_*^{n-1}\right]\;.
\ea

From \eq{rho-1} to \eq{pperp-1}, it is then easy to see that the isotropic limit can be achieved if $\L_n$ is large enough~\footnote{Since the UV $Z_n$ symmetry is approximately good  if $\Phi_0\gg m$, then the isotropic limit $\L_n=\Big(\l_n\frac{\Phi_0^2}{m^2}\Big)^{2\over n-2}  \Big(\frac{M_{Pl}}{ \sqrt{4\pi }\Phi_0}\Big)^2\gg 1$ is easier to achieve. It is noticed that a large derivative of $ \sigma_*$ may balance the large $\Lambda_n$ term, but~here we only consider the simplest case.} so that the spatial kinetic term can be neglected in comparison to the other terms~\footnote{Naively, the~comparison should also take into account the spatial profile of $A(x)$, $B(x)$ and $\sigma_*(x)$. Here, we assume that $\L_n$ is large enough as argued in the previous footnote so that the variation of the spatial profiles of $A$, $B$ and $\sigma_*$ will not affect the result. This of course can be justified after solving the TOV configurations.}. Thus, the~resultant approximate energy density and isotropic pressure then become 
\bea
 \rho&=& {m^2 M_{Pl}^2  \over 4\pi\L_n}\Big[ {1\over 2}({\Omega^2 \over B}+1)\sigma_*^2 +  {1\over n}\sigma_*^n \Big], \\
 p&=& p_{\perp}={m^2 M_{Pl}^2 \over 4\pi\L_n}\Big[ {1\over 2}({\Omega^2 \over B}-1)\sigma_*^2 -  {1\over n}\sigma_*^n \Big]. 
\eea

Moreover, in~this isotropic limit, both $\rho$ and $p$ depend only on $\sigma_*$ and not on its spatial derivative, and~the first term of \eq{reduced key} is suppressed with respect to the second term\footnote{The same assumption is adopted as in the previous footnote to justify the suppression.} so that it can be solved for $\sigma_*$ to yield 
\be\label{phin-s}
\sigma_*^{n-2}=\frac{\Omega^2}{B}-1 \;.
\ee

Using \eq{phin-s}, the~energy density and pressure can be expressed as a pure function of $\sigma_*$,
\bea
 \rho&=& \rho_{n,0} \Big[ ({1\over 2}+{1\over n})\sigma_*^n + \sigma_*^2 \Big], \label{rho-eos} \\
 p&=& \rho_{n,0} ({1\over 2}-{1\over n})\sigma_*^n ,\label{p-eos}
\eea
where the overall energy density scale is given by 
\be
\rho_{n,0}= {m^2 M_{Pl}^2  \over 4\pi\L_n}= {4\pi m^2 \Phi_0^2 }({m \over{\sqrt \l_n \Phi_0}  })^{\frac4{n-2}}\;.
\ee

Note that \eq{rho-eos} and \eq{p-eos} are already forming a parametric EoS for the dark matter model in the isotropic limit. However, we can further eliminate the parametric function $\sigma_*(r)$ to yield a compact form of EoS as following (in the units of $c=1$ and $\hbar=1$):
\be\label{phin}
{\rho\over \rho_{n,0}}={n+2\over n-2} 
\Big({p\over \rho_{n,0}}\Big) + \Big({2n \over n-2}  \; {p\over \rho_{n,0}}\Big)^{2\over n}.
\ee

One can then adopt this EoS to solve the TOV configurations for the dark or hybrid stars. Note~that, for~$n=4$, the~above EoS reduces to the known result given in~\cite{Colpi:1986ye}, namely,
\be\label{phi4colpi}
{\rho\over \rho_{4,0}}=3\Big({p\over \rho_{4,0}}\Big) +2 \sqrt{p\over \rho_{4,0}}\,.
\ee

For the other $n$'s, the~resultant EoSs are new and not considered before in the literature. It is interesting to see that, as~$n$ goes higher, the~EoS becomes stiffer, e.g.,~as $n\longrightarrow \infty$, the~EoS goes to $p=\rho$, i.e.,~the causality~limit.  

Even though we have arrived at the EoS \eq{phi4colpi}, it is written in the unit of $\rho_0$. For~the convenience of later use when solving TOV configurations, we will choose the following astrophysical units associated with the solar mass $M_{\odot}$:
\be\label{astrophy}
r_{\odot}=G_N M_{\odot}/c^2,\quad \rho_{\odot}=M_{\odot}/r_{\odot}^3, \quad p_{\odot}= c^2 \rho_{\odot}.
\ee 
  
Then, in~terms of these units, the~EoS \eq{phi4colpi} will turn into the following form~\cite{Zhang:2020pfh}:
\be\label{DEoS}
{\rho\over \rho_{\odot}}=3 \; \Big({p\over p_{\odot}}\Big)+{\cal B}_4\; \sqrt{p\over p_{\odot}}\,
\ee
with ${\cal B}_4:= {0.08 \over \sqrt{\lambda_4}}({m\over \textrm{GeV}})^2$ a free  parameter.  Similarly, the~EoS \eq{phin} can be turn into 
\be\label{EoS_phin}
{\rho\over \rho_{\odot}}={n+2\over n-2}\; \Big({p\over p_{\odot}}\Big)+{\cal B}_n\; \Big({p\over p_{\odot}}\Big)^{2\over n}
\ee
where ${\cal B}_n:=({2n \over n-2}  )^{2\over n} ({\rho_{n,0} \over p_{\odot}})^{1-{2\over n}}$ is again a free parameter related to $m$ and $\lambda_n$. 
   
 \begin{figure}[htbp] 
\centering
{\includegraphics[width=7.5cm]{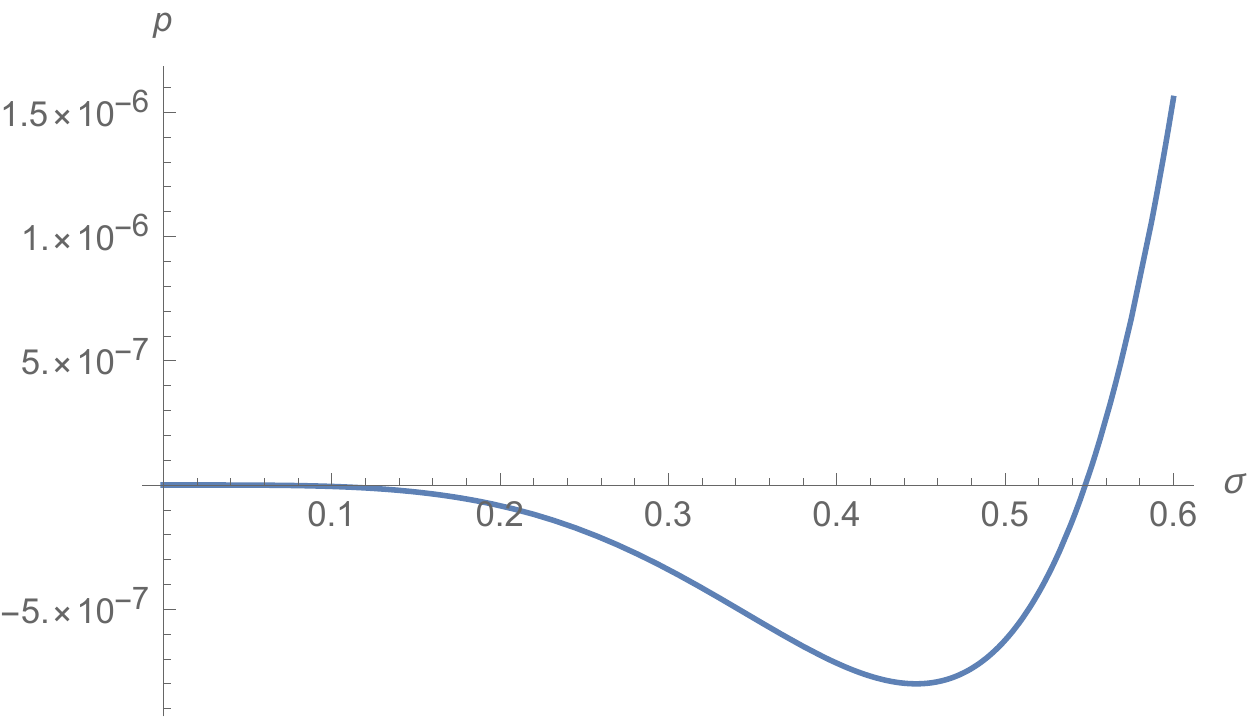} }%
\quad 
{\includegraphics[height=0.30\textwidth]{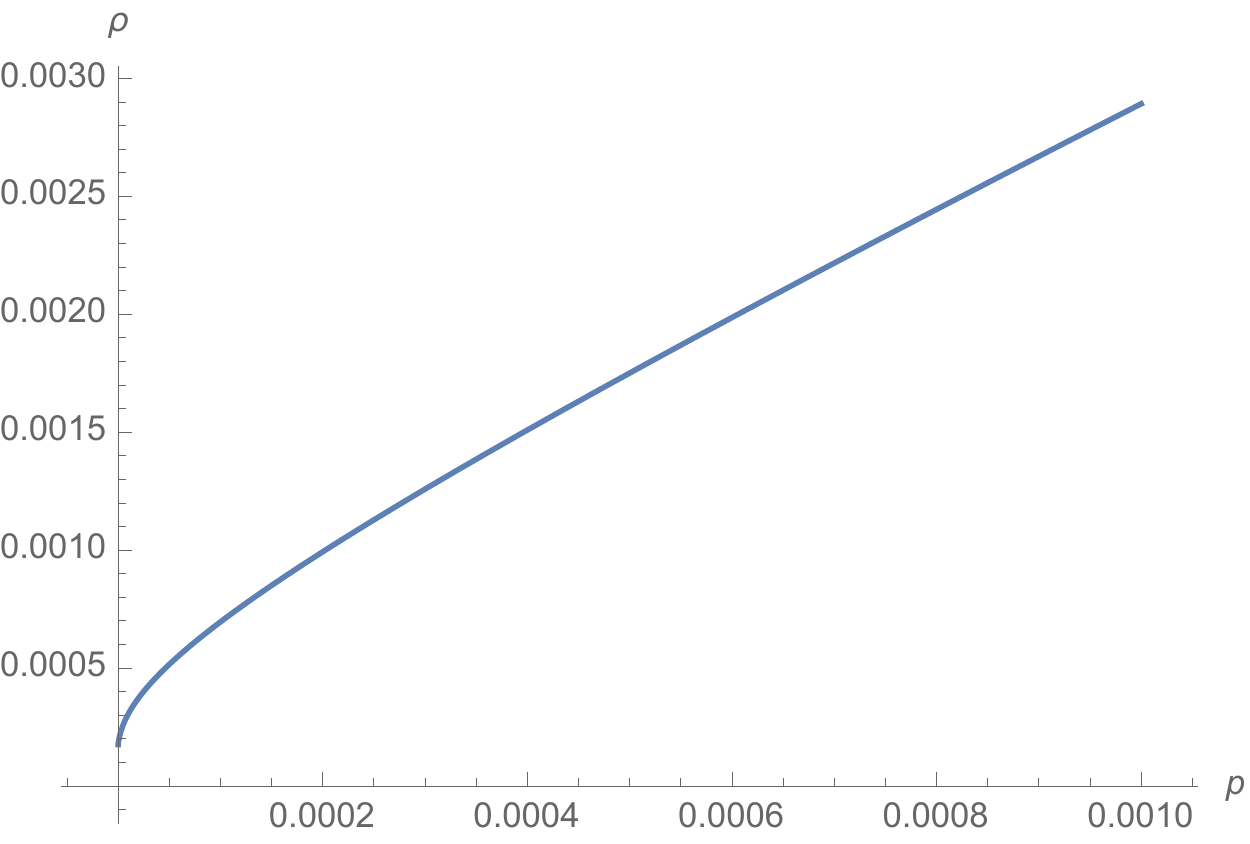} }%
\caption{An example of EoS given by \eq{EoS46-1} and \eq{EoS46-2} with ${\cal B}=0.0006$ and ${\beta}=0.1$. Left: it shows the region of $p<0$ which should be excluded when solving TOV configurations of dark stars although it may yield gravastar configurations. Right: the plot of EoS for the phsyical region. The~unit of $p$ is $p_{\odot}$, and~the unit of $\rho$ is $\rho_{\odot}$, and~$\sigma$ is~dimensionless.  }
\label{prhosigma}
\end{figure}  
   
The above method of extracting the EoS in the isotropic limit from a given  scalar field theory can be applied to the model with more general potentials other than \eq{Vn}.  For~example, for~the model with the following potential,
\be
V(\phi)=\frac12 m^2|\phi|^2-\frac14 \l_4 |\phi|^4+\frac16 {\l_6 \over \Phi_0^2} |\phi|^6\,.
\ee

In this case, we have two scaling parameters $\L_4$ and $\L_6$ as defined in \eq{Lambda_n}. However, it is easier to parameterize the EoS in the isotropic limit by the following two parameters: $\L:=\sqrt{\L_6}$ and $\beta:={\L_4 \over 4 \L}$ so that the extracted EoS takes the following form in terms of the astrophysical units of \eq{astrophy}:
\ba\label{EoS46-1}
{\rho\over \rho_{\odot}}&=&{\cal B} \left(\frac23 \sigma^6-3  \beta \sigma^4+\sigma ^2\right) \,
\\ \label{EoS46-2}
{p\over p_{\odot}}&=&{\cal B} \left(\frac13 \sigma^6-  \beta \sigma^4\right)
\ea
where ${\cal B}:=\frac{m^2 M^2_{pl}}{4 \pi \Lambda \rho_{\odot}}$.  

Some comments about the above EoS are in order: (i) one can see ${\cal B}>0$ but $\beta$ can be either positive or negative. When $\beta=0$, it reduces to the case of pure $\phi^6$ model with ${\cal B}_6=(3{\cal B}^2)^{1/3}$; (ii)~We~should require both $\rho \ge 0$ for positive energy condition and $p\ge 0$ for not considering the gravastars, thus the corresponding physical range of $\sigma$ should be chosen carefully. In~particular, when $\beta>0$, one should require $\sigma \ge \sqrt{3\beta}$ to keep $p\ge 0$. An~example for this case is shown in Figure~\ref{prhosigma}. On~the other hand, both~$\rho$ and $p$ are positive for all ranges of $\sigma$ when $\beta<0$.

Below, we will study the TOV configurations of dark and hybrid stars made of nuclear and dark matters based on the above EoSs of bosonic SIDM model in the isotropic limit. In~particular, we will focus more on the hybrid stars with saddle instability which causes more marginal parameter space to constrain the dark matter models by the observed LIGO/Virgo gravitational wave events. In~this way, we demonstrate how to constrain the particle physics models of dark matter by the gravitational wave astronomical~observations.

\section{BTM Criteria and Saddle Instability for Hybrid~Stars} \label{sec 3}
Neutron stars and some other compact stars are relativistic objects that their structure should be analyzed using general relativity.
TOV equations~\cite{TOV-1,TOV-2,Ott13} can be applied for these kinds of calculations, which are 
derived from the Einstein equations and the conservation of the energy--stress tensor, 
assuming zero space velocity, spherical symmetry, and~an ideal fluid~model.

In this paper, we will mainly consider the hybrid stars by solving the following TOV configurations with multiple component fluids inside the star~\cite{ Mukhopadhyay:2016dsg, Rezaei:2018cuk}:
\be
{dp_I \over dr}=-(\rho_I + p_I) {d \phi\over dr},\quad  {d m_I\over dr} =4\pi r^2 \rho_I, \quad {d \phi \over dr} ={m +4\pi r^3 p \over r(r-2m)},
\ee
where the index $I$ labels the fluid components, and~the total pressure and energy density are given by $p:=\sum_I p_I$  and $\rho:=\sum_I \rho_I$, respectively; and $m(r)=\sum_I m_I(r)$ is the mass profile inside the star, and~the Newton potential $\phi:={1\over 2} \ln(-g_{tt})$ with $g_{tt}$ the $tt$-component of the metric. In~this paper, we~will consider the cases with only two-component hybrid stars made of nuclear matter and dark~matter. 

The resulting configurations of hybrid stars depend on if there are strong interactions between component fluids. If~there are strong interactions, it is expected to form a domain wall between phases of different fluids. Following our previous consideration in~\cite{Zhang:2020pfh}, we refer the hybrid stars with neutron core and dark matter crust as scenario I, and~the ones with dark matter core and neutron crust as scenario II. On~the other hand, if~there is almost no interaction among component fluids, then all components prevail inside the star and mix together right from the core. We refer to this case as scenario III~\cite{Zhang:2020pfh}. To~solve the TOV equations, we need to provide the core pressure as the initial condition, and~then, by~changing the initial pressure, we will obtain different TOV configurations to plot the mass--radius relation. For~scenarios I and II, one only needs to provide one initial condition for the core pressure since only one kind of fluid dominating the core, but~needs two for scenario~III.

Not every TOV configuration is stable, and~one needs to judge its stability by either solving the Sturm--Liouville eigenmodes of radial oscillation (another method of judging the equilibrium configuration is used in~\cite{Henriques:1990xg, Valdez-Alvarado:2020vqa, DiGiovanni:2020frc}) or by some equivalent empirical criteria. One set of such criteria for single component fluid is the so-called BTM (Bardeen--Thorne--Meltzer) criteria~\cite{BTM}, which~states that, in~the direction of increasing core pressure along the mass--radius curve, whenever an extremum is passed, one stable mode becomes unstable if the curve bends counterclockwise. Contrarily, one~unstable mode becomes stable if it rotates clockwise. These criteria were originally formulated by starting from the stable planet configuration with low enough core pressure. It, however, will~cause some trouble if one does not start from such kind of planet configuration when solving the TOV equations. Thus, in~practicality, it is useful to formulate the reverse BTM criterion by traveling the mass--radius curve in the direction of decreasing core pressure, and~to require the consistency with the BTM ones. This then leads to following { (Reverse)
 BTM stability criteria:} { Traveling the mass--radius curve along either directions of increasing or decreasing core pressure, whenever passing an extremum, one~stable mode becomes unstable if the cure bends counterclockwise; otherwise, one~unstable mode becomes stable.}

\begin{figure}[htbp] 
\centering
{\includegraphics[height=0.44\textwidth]{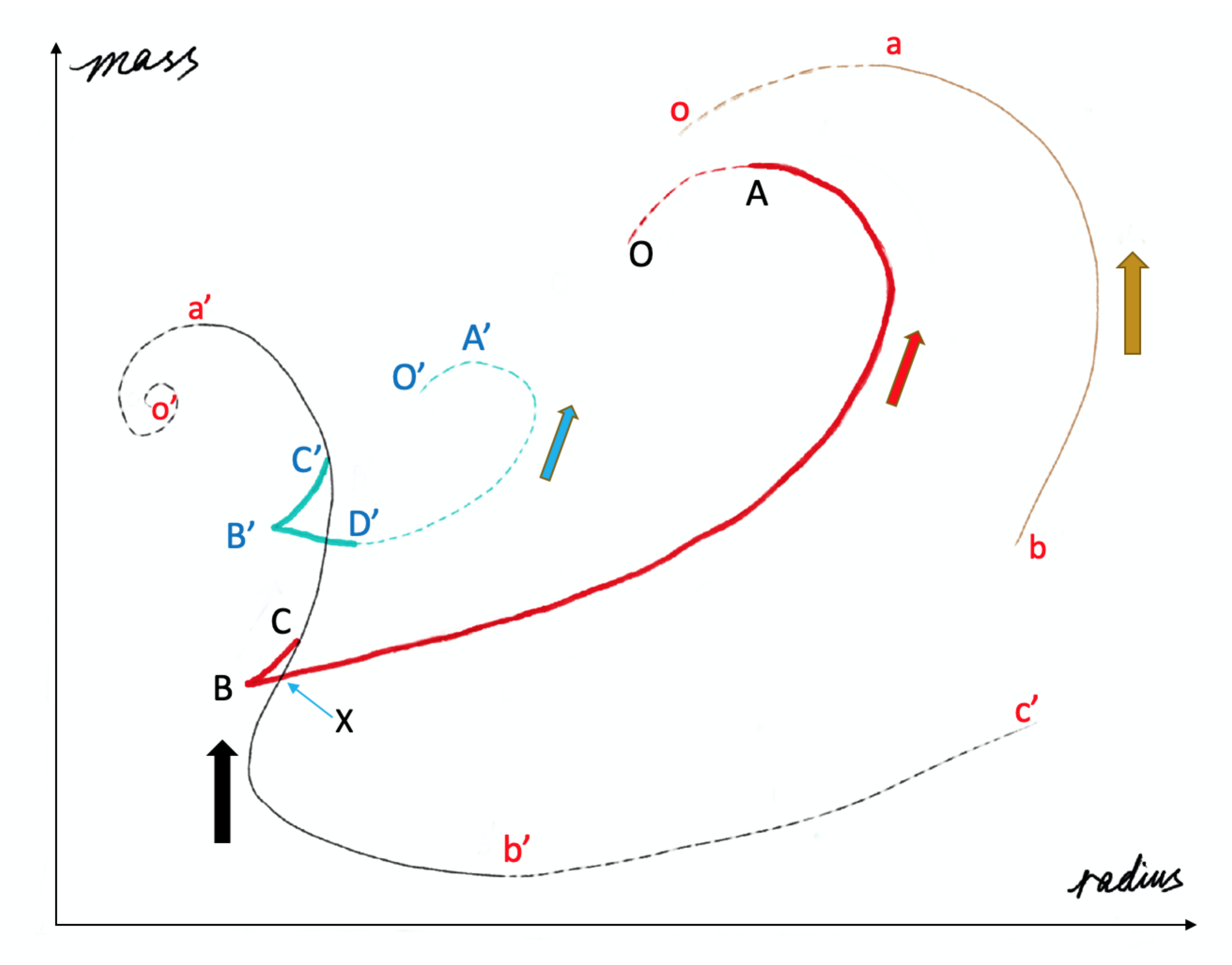}
\quad
\includegraphics[height=0.26\textwidth]{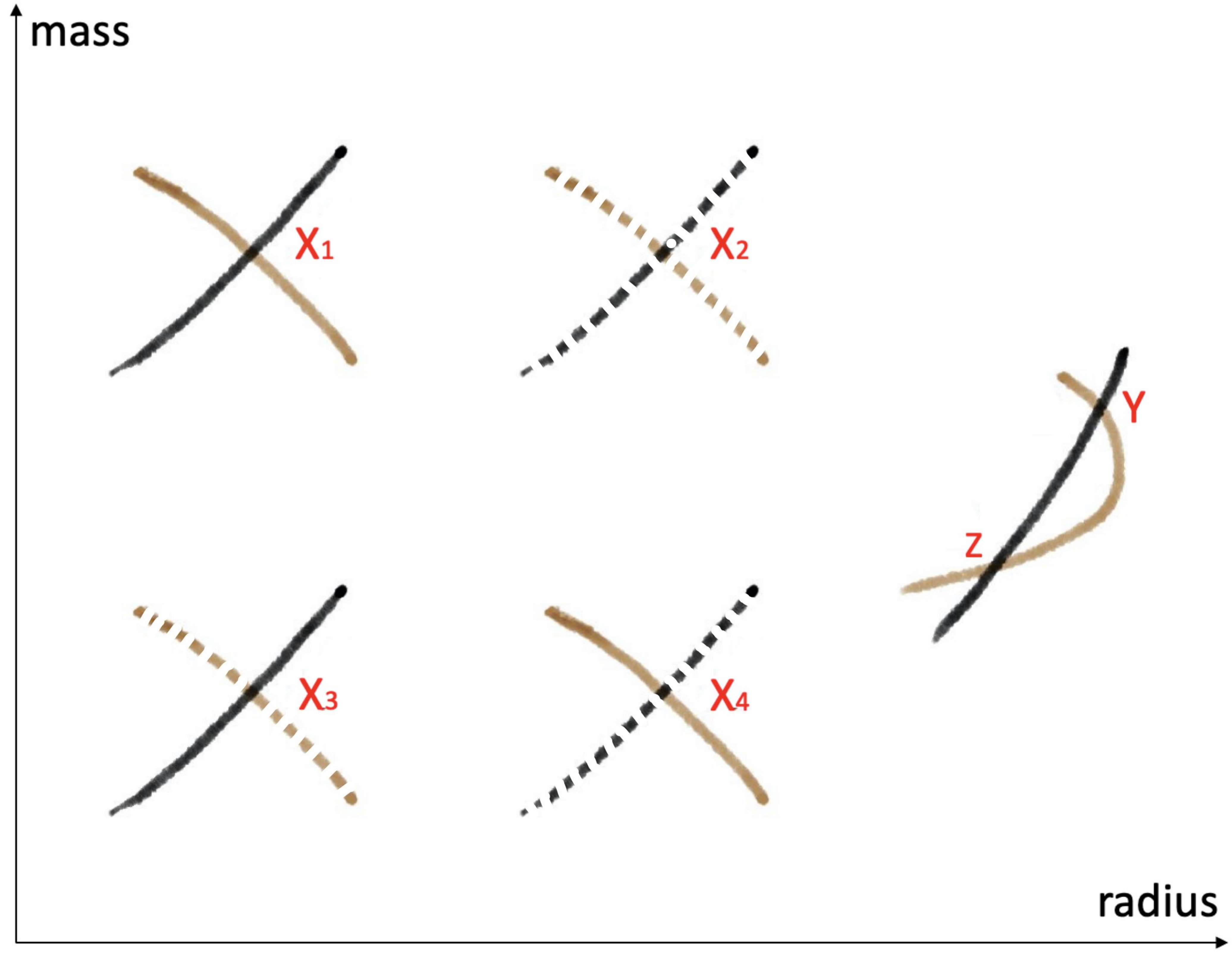}
}
\caption{The various curves and intersections of mass--radius relations used to demonstrate the (reverse) BTM criteria in the main text. Solid lines indicate stable parts, while the dotted lines are for unstable parts. Left panel: Various mass--radius curves. The~thick arrows indicate the directions of increasing core pressures for each branch. Right panel: Summary of the situations of intersections. $X_1$~is a stable point where both the neutron branch (black) and the dark branch (brown) are stable.  $X_2$ is unstable since both branches are unstable.  $X_3$ and $X_4$ show the saddle instability that one of the two branches is unstable. In~some cases, the~two branches may cross twice like points $Y$ and $Z$; then, at~ least one of them is a fake~intersection. }
\label{BTM}
\end{figure}

By applying the (reverse) BTM criteria, one can make sure that, for~the unstable regions on the mass--radius curve, not even one starts from a known stable region. However, one can only pin down the stable regions if one starts from a stable region, or~knows how many unstable modes there are. For~example, for~the curve $O'A'D'B'C'$ on the left panel of Figure~\ref{BTM}, one can apply the (reverse) BTM criteria criteria to make sure $O'A'D'$ is unstable regardless if $D'B'C'$ is stable or not, even if seems impossible to apply the original BTM criterion as there is no asymptotic stable region. On~the other hand, for~the curve $OABC$ in Figure~\ref{BTM}, one cannot be sure if $ABC$ is stable even if one can make sure that $OA$ is unstable by applying the (reverse) BTM~criteria. 

Importantly, the~above criteria is only checked rigorously for the stars with single component fluid. One should be careful when applying the (reverse) BTM criteria to judge the stability of hybrid stars. For~the hybrid stars of scenarios I and II, there is only one dominant fluid component in each region, one will then expect that the (reverse) BTM criteria should still work without the need of much modification; see~\cite{Kain:2020zjs} for the recent~discussion. 

On the other hand, for~the hybrid stars of scenario III, one needs to set the core pressures for both fluid components, say one for nuclear matter and one for dark matter.  Thus, one needs to plot two kinds of mass--radius curves, namely, one by fixing the neutron core pressure but tuning the dark matter core pressure, and~the other by fixing the dark matter core pressure but tuning the neutron core pressure. Let us call the former the dark branch and the latter the neutron branch. In~the left panel of Figure~\ref{BTM}, we show some representatives of both branches, e.g.,~the curve $o'a'b'c'$ is the neutron branch, and~the three curves in the same figure are the dark branches\footnote{Neutron branches could also look like the two midlle curves $O'A'D'B'C'$ and $OABC$, but~with oppisite pressure directions, like the solid curves in {Figures} \ref{phi4}--\ref{phisix}.
}. For~any intersection point of dark and neutron branches, e.g.,~point $X_1$ to $X_4$ on the right panel of Figure~\ref{BTM}, there could be saddle instability\footnote{For a single-component star, the~stability is solely determined by altering the core pressure. When the core pressure has some perturbation, if~the radial oscillation is stable, then the star is stable. However, for~a two-component star, the~core pressure is a sum of two partial pressures of each component. Then, a~necessary condition for such a star being stable is that it is stable when changing either of the partial pressure while keeping the other fixed. It might happen that a star is stable when changing the neutron (or dark matter) pressure but unstable when changing the other partial pressure. This will still lead to an unstable star and we call it the saddle instability. } if one of the branches is unstable at this intersection point.  The~difficulty is how to judge the stability for each branch. Without~rigorous study of the Sturm--Liouville eigenmodes of radial oscillation, it is hard to answer this question. In~this work, we assume that the (reverse) BTM criteria still work for each dark or neutron branch, and~apply the criteria to find out the saddle unstable regions. Furthermore, we shall also assume that the cusp points such as $B$ and $B'$ ($B'$ is not even an extremum) on the left panel of Figure~\ref{BTM} will not induce the change of stability of any radial oscillation eigenmode. Otherwise, there could be more complications~\cite{Alford:2017vca}.

Even with the above assumptions holding when traveling along the mass--radius curve, to~judge the stability with (reverse) BTM criteria, it is better to start from some known stable region. One such region is the stable part of the pure neutron star curve, and~the other is the stable part of the pure dark star curve. We then need to further assume that the small doping with the other component will not change the stability; it then implies that the nearby regions of the stable part of single component fluid stars are also stable. For~example, if~the curve $o'a'b'c'$ is a nearby curve around the pure neutron star, and~its stable part (the solid line $a'b'$) is inherited from the stable part of the pure neutron star, then one can infer the unstable part (the dashed line $o'a'$ and $b'c'$) by applying the (reverse) BTM criteria. Furthermore, the~nearby regions such as $C'B'D'$ and $CBX$ should also be stable. The~reason is as follows: these two regions branch out from an almost pure neutron star curve, and~are at the ends of dark branches, this means that the core pressures of the dark matter component are negligible. Thus, these two parts are indeed almost pure stable neutron stars. We can then infer that: (1) $D'A'O'$ is unstable because $B'D'A'O'$ bends counterclockwise when passing around local extrema $D'$ and $A'$; and (2)$XA$ is stable as there is no local extremum on it. (3) $A'O'$ is unstable because it bends counterclockwise when passing through the local maximum $A$.
 
Based on the above discussions, we can then apply the reverse BTM criteria to the intersection points of the neutron and dark branches, such as the point A on the right panel of Figure~\ref{BTM} to judge if the intersection point is saddle stable or not. Both curves crossing at $A$ being stable is a necessary condition for $A$ to be stable. In~addition, there are some fake intersection points such as point $X$ for the case when one curve intersects with the other curve more than once. Then, only the end point/branch point is real intersection and others are fake. We have summarized these situations in the right panel of Figure~\ref{BTM}. We will apply the lessons from the above discussions to the case studies in the next section to judge the stability of hybrid stars, especially the saddle instability of scenario~III.

\section{Dark Star and and Hybrid Star~Interpretations} \label{sec 4}

In this section, we consider the dark stars and hybrid stars of all three scenarios based on the EoSs extracted from the bosonic SIDM models discussed in Section~\ref{sec 2} in the isotropic limit. For~the EoS of the nuclear matter, the~standard  phenomenological neutron EoS SLy4~\cite{Douchin:2001sv,SLy4} is employed, which has a maximum mass of 2.05 $M_{\odot}$, and~the radius at 1.4 $M_{\odot}$ is about 12km. This EoS is offered as discrete sets of pressure and energy density, which is convenient to be dealt with numerically.
 The main purpose is to see if some of the dark or hybrid stars can reach the mass of 2.6 $M_{\odot}$ or so to explain the smaller companion compact object of GW190814. It is easy to see that this purpose can be easily achieved for the dark stars and hybrid stars of scenarios I and II by choosing appropriate parameters of SIDM. On~the other hand, it is more difficult for the hybrid stars of scenario III to reach such a mass because of the saddle instability. It is interesting to see that the astronomical observations of gravitational waves can rule out some theoretical scenarios of hybrid stars and the associated dark matter models. Thus, we will focus more on scenario III in the later discussions of this~section.

Now, we present the mass--radius relations case by~case.

 \begin{figure}[htbp] 
    \centering
{\includegraphics[height=0.55\textwidth]{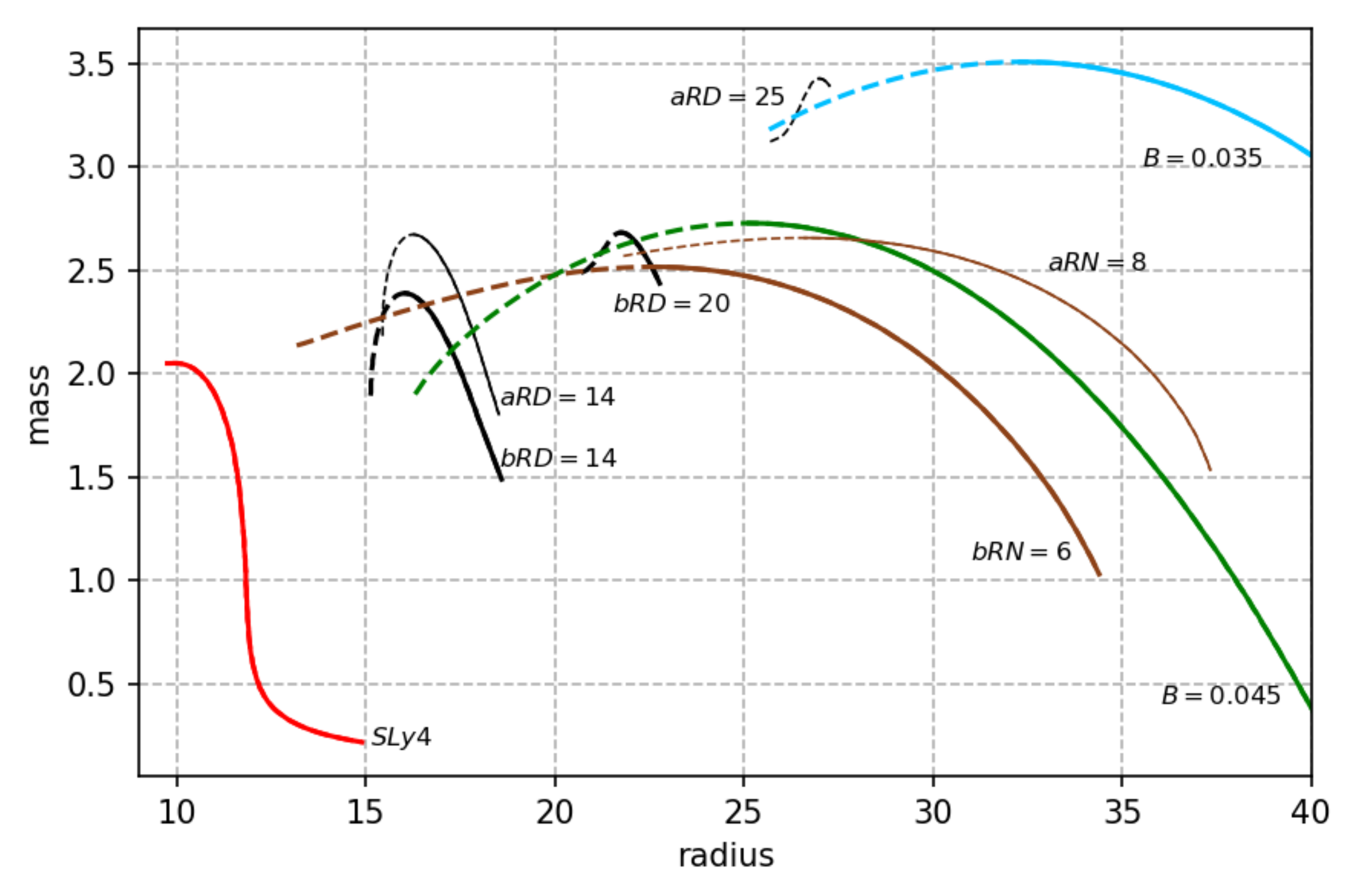}}
\quad
{\includegraphics[height=0.55\textwidth]{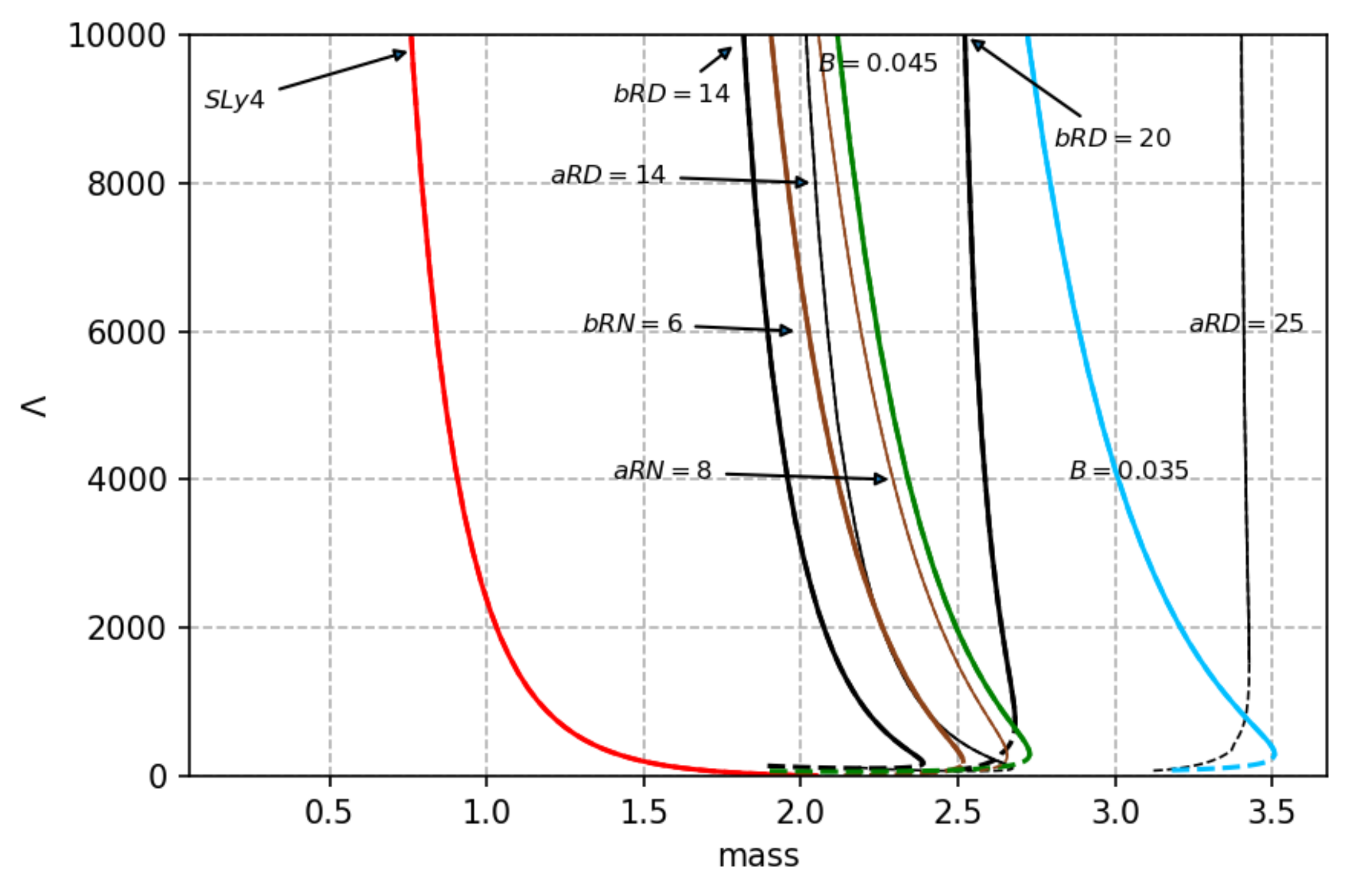}}%
\caption{{Mass-Radius} (up) and TLN-Mass (down) relations for the hybrid stars of scenarios I and II. The~red line stands for the pure neutron stars with SLy4 EoS,  while the green and blue lines stand for the pure dark stars of EoS given by \eq{DEoS} with ${\cal B}_4 =0.035$ and $0.045$, respectively. The~other lines represent hybrid stars of neutron and dark matters associated with the above EoSs. For~${\cal B}_4$ = 0.035, they are labelled by {\bf aRN} = $r_W$ (thin brown lines) for hybrid stars of neutron core, and~{\bf aRD} = $r_W$ (thin~ black lines), where $r_w$ is the core radius. Similarly, for~${\cal B}_4 =0.045$, they are by {\bf bRN} = $r_W$ (thick brown lines) and {\bf bRD} = $r_W$ (thick black lines). For~example, {\bf bRN} = 6 labels the hybrid stars of $r_W=6 $ km. The~unstable configurations are denoted by the dashed lines. The~bottom panel is presented to show the orders of the corresponding TLN, and~we find that 2.6 $M_{\odot}$ stars are well below 4000, which are negligible after averaging with a 23 $M_{\odot}$ black hole for GW190814. The~unit of mass is $M_{\odot}$, and~the unit of radius is km, and~TLN $\Lambda$ is~dimensionless.}
\label{rdrn}
\end{figure}

{Bosonic
 $\phi^4$ model. } 
We first consider dark and hybrid stars based on the $\phi^4$ EoS, i.e.,~\eq{DEoS}, which is extracted from the $\phi^4$ dark SIDM in the isotropic limit. For~scenarios I and II, i.e.,~assuming interactions exist between neutron and dark matter, it is easy to form a 2.6 $M_{\odot}$ star, as~shown in Figure~\ref{rdrn}, in~which we also show the TLN-mass relations for one's reference. Pure neutron stars of SLy4 EoS are marked in red, and~pure dark stars with ${\cal B}_4=0.035$ and $0.045$ are marked in blue and green, respectively. The~other lines represent hybrid stars of neutron and dark matter with these EoS.  When~${\cal B}_4 =0.035$, they are labelled  by {\bf aRN}  = $r_W$ (brown) for neutron core case and by {\bf aRD} = $r_W$ (black) for dark matter core case. Here, $r_W$ stands for the core radius. Similarly, when ${\cal B}_4 = 0.045$, they~are labelled by {\bf bRN} = $r_W$ (brown) and {\bf bRD} =$r_W$ (black)). For~example, {\bf bRN} = 6 means the radius of neutron core is  $6\textrm{km}$, and~${\cal B}_4=0.045$. The~unstable configurations are denoted by dashed~lines.

We find that this dark star model can cover any mass range by adjusting the free parameter ${\cal B}_4$. In~particular, for~${\cal B}_4 \leq 0.047$, the~maximal mass exceeds $2.6$ $M_{\odot}$. The~maximal mass grows without an upper limit as the value of ${\cal B}_4$ drops, and, for~example, when ${\cal B}_4 =0.035$, we have $3.5$ $M_{\odot}$. Since~for scenarios I and II the 2.6 $M_{\odot}$ hybrid stars can always be achieved as long as the pure dark star has a maximal mass higher than 2.6 $M_{\odot}$, in~the following, we only concentrate on the more nontrivial scenario III, which is more realistic since it is assumed that there is no interaction between SIDM and baryonic~matter.

\begin{figure}[htbp] 
\centering
\includegraphics[height=0.5\textwidth]{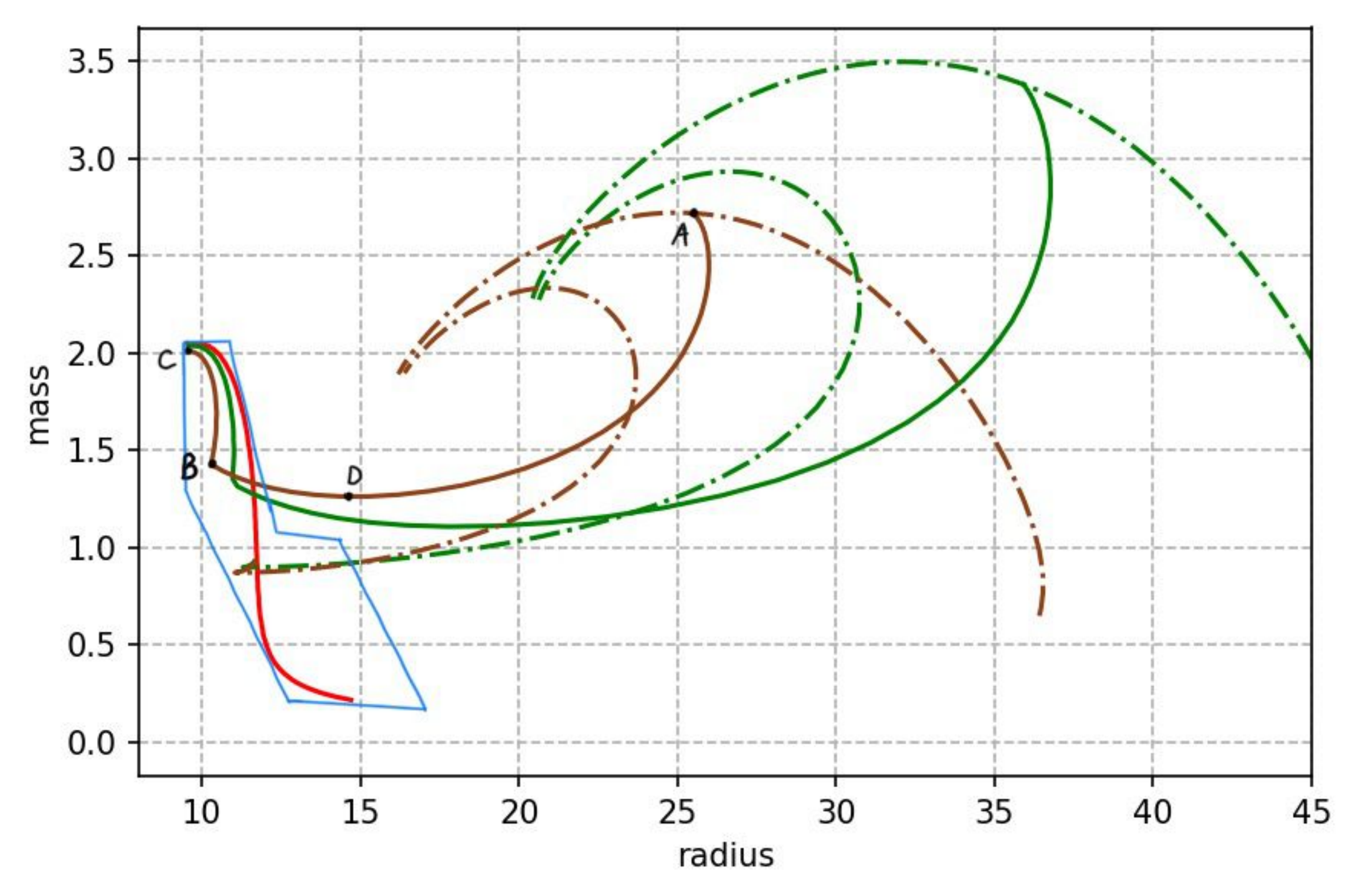}   
\caption{Mass--radius relation for the hybrid stars of scenario III, with~SLy4 EoS for neutron and $\phi^4$ model EoS given by \eq{DEoS} for dark matter. The~pure SLy4 neutron stars are marked in red for one's reference. The~hybrid star configurations are shown in brown for ${\cal B}_4=0.045$, and~in green for ${\cal B}_4=0.035$. Solid lines represent the neutron branches, and~dash-dotted lines the dark branches. The~blue-circled regions roughly indicate the stable hybrid star configurations for ${\cal B}_4=0.045$. Saddle~instabilities can be checked at each intersection point of solid and dash-dotted lines with the same color. For~example, point A is unstable since the solid brown line on it is unstable, though~the dash-dotted line on it is stable. The~unit of mass is $M_{\odot}$, and~the unit of radius is~km.}
\label{phi4}
\end{figure}

For scenario III, by~applying the (reverse) BTM stability criteria to the resultant mass--radius relation shown in Figure~\ref{phi4}, no stable stars around 2.6 $M_{\odot}$ can be formed. 
The pure SLy4 neutron stars are marked in red as the reference configurations. For~the hybrid stars, ${\cal B}_4=0.045$ case is denoted by the brown lines,  and~$0.035$ case by the green lines. Now, we need the core pressures for both dark and nuclear matters to solve the TOV configurations. By~tuning one of the core pressures and fixing the other, we can obtain the so-called dark branch and neutron branch as discussed in Section~\ref{sec 3}. In~ Figure~\ref{phi4}, the~dark branches are denoted by the dash-dotted lines and the neutron branches by the solid~lines. 

To apply the method in Section~\ref{sec 3} to check the saddle (in)stability, we can apply the (reverse) BTM criteria to both neutron and dark branches and then determine the saddle stability for their intersection points. As~discussed in Section~\ref{sec 3}, we also need to assume that the nearby regions of the pure stable star configurations are stable when applying (reverse) BTM criteria.  After~checking this way for most of the intersections in Figure~\ref{phi4}, we can determine the stable regions and unstable regions. For~example, the~point $A$ is unstable because the line $BDA$ turns counterclockwise at point $D$ towards $A$, which~makes the $DA$ part unstable. It turns out that the stable regions are small and confined near the pure neutron star configurations\footnote{It is interesting to notice that the region near pure dark star (like point A) is unstable. This should be understood that, although~the dark matter components are stable, the~neutron parts are unstable, which makes the total configuration unstable.  }, which for ${\cal B}_4=0.045$ is roughly indicated by the regions inside the highlighted blue closed line. { From
 the above analysis, we can conclude that, for~the bosonic $\phi^4$ model, however the parameter ${\cal B}_4$ varies, the~stable region cannot yield the mass of the star more than 2.1 $M_{\odot}$, which is just the maximal mass of pure neutron star associated with SLy4 EoS.} This could be due to the fact that the EoS associated with $\phi^4$ model is not stiff enough when compared with SLY4. This can also be seen from the fact that the radius of a dark star of 2.6 $M_{\odot}$ is at least 25 $\textrm{km}$, far larger than the standard neutron star's radius around 11 $\textrm{km}$. 

\begin{figure}[htbp] 
    \centering
\includegraphics[height=0.5\textwidth]{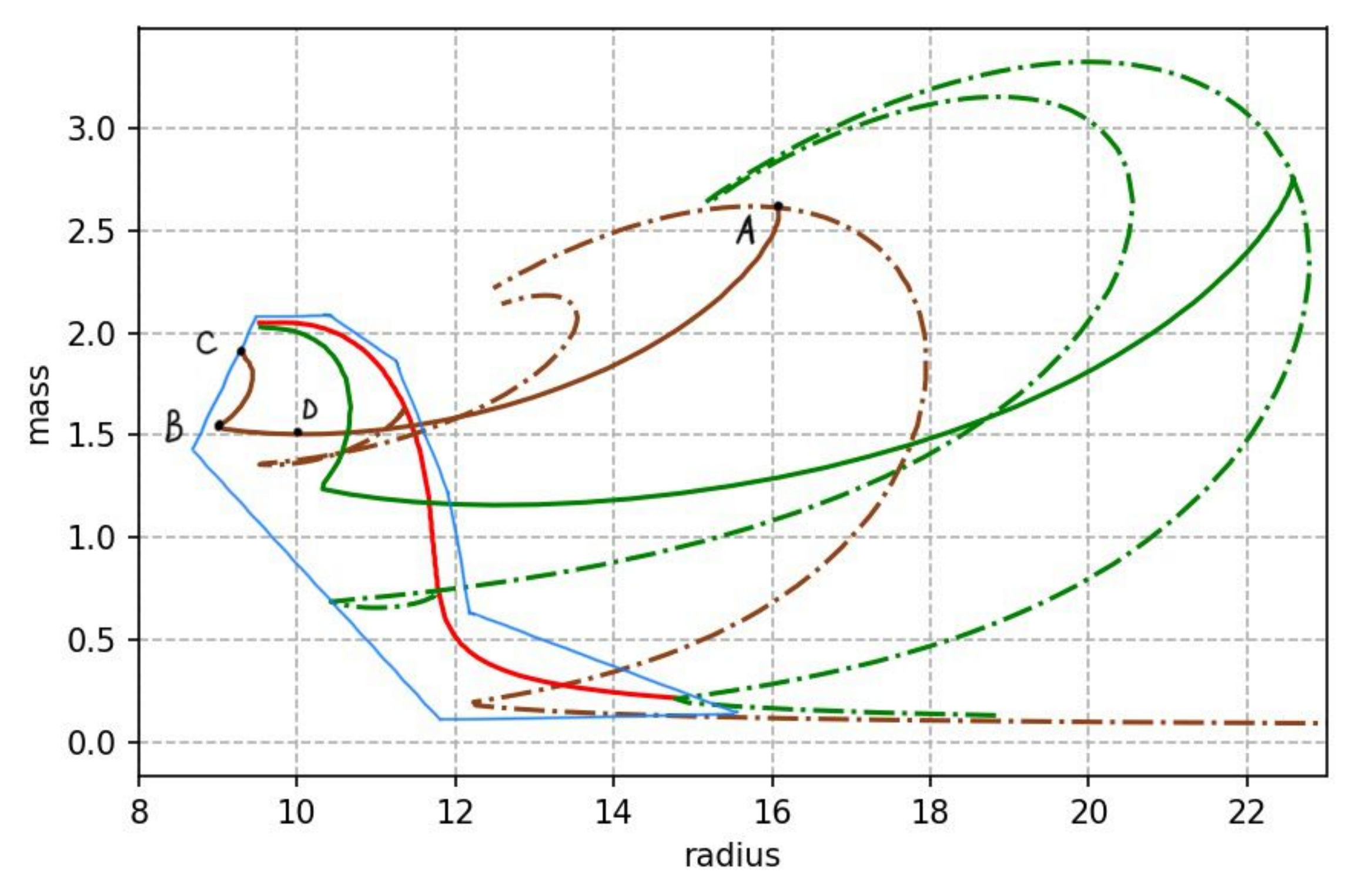}
\caption{Mass--radius relation for the hybrid stars of scenario III, with~SLy4 EoS for neutron and $\phi^6$ model EoS given by \eq{EoS_phin} with $n=6$ for dark matter. The~line styles are the same as in Figure~\ref{phi4} but with the brown lines for ${\cal B}_6=0.011$ and the green lines for ${\cal B}_6=0.008$. The~blue-circled regions roughly indicate the stable hybrid star configurations for ${\cal B}_6=0.011$. Saddle instabilities can be checked at each intersection point of solid and dash-dotted lines with the same color. The~unit of mass is $M_{\odot}$, and~the unit of radius is~km.}
\label{phi6}
\end{figure}

\begin{figure}[htbp] 
    \centering
\includegraphics[height=0.5\textwidth]{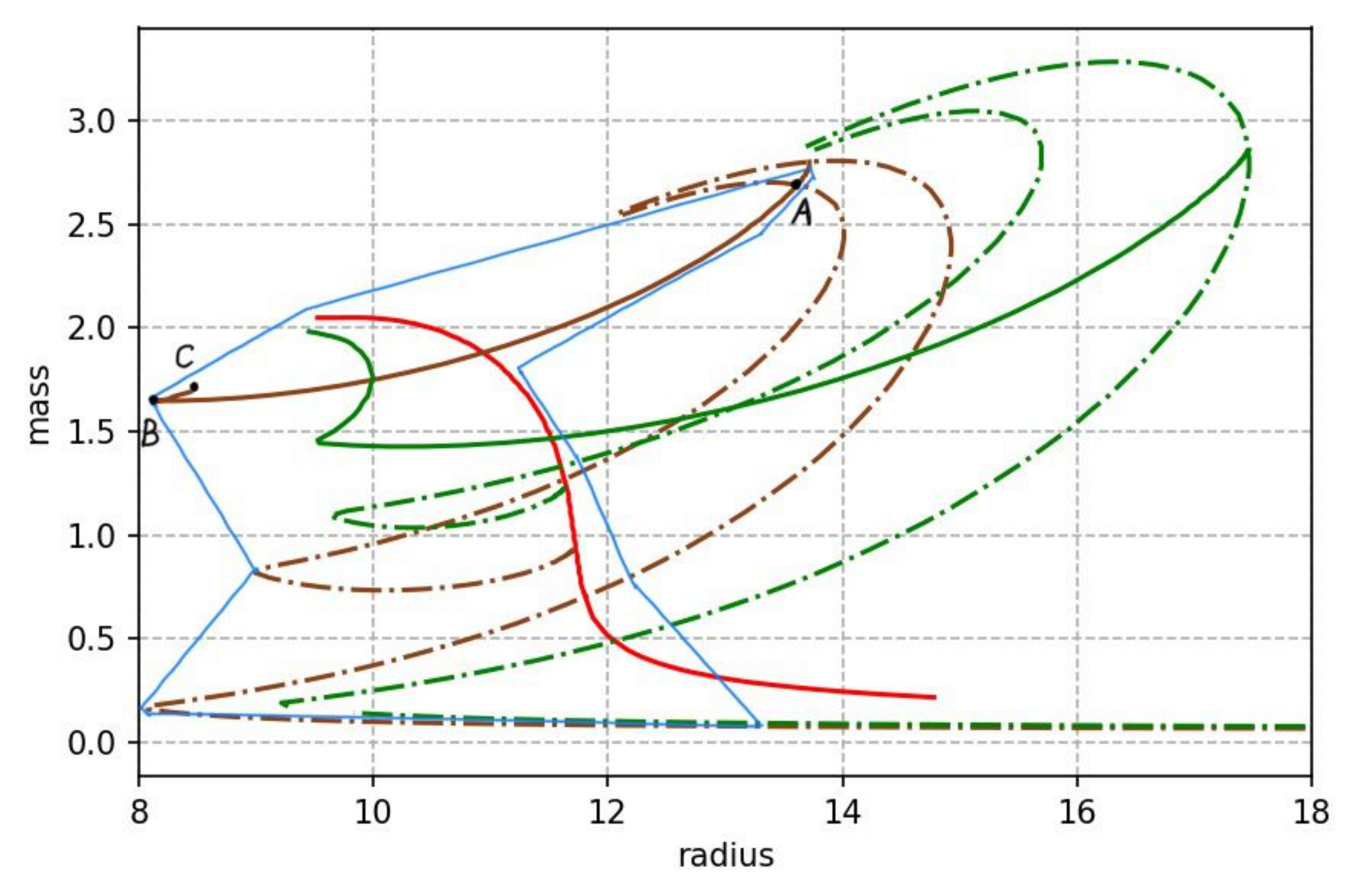}
\caption{Mass--radius relation for the hybrid stars of scenario III, with~SLy4 EoS for neutron and $\phi^{10}$ model EoS given by \eq{EoS_phin} with $n=10$ for dark matter. The~line styles are the same as in Figure~\ref{phi4} but with the brown lines for ${\cal B}_{10}=0.0036$ and the green lines for ${\cal B}_{10}=0.0028$. The~blue-circled regions roughly indicate the stable hybrid star configurations for ${\cal B}_{10}=0.0036$. Saddle instabilities can be checked at each intersection point of solid and dash-dotted lines with the same color. The~unit of mass is $M_{\odot}$, and~the unit of radius is~km.}
\label{phi10}
\end{figure}

{ Bosonic $\phi^n$ model. }
From the expression of $\phi^n$ EoS, i.e.,~\eq{EoS_phin}, we know it becomes stiffer as $n$ grows. This makes it possible to have more massive stable hybrid star configurations. Indeed, from~Figures \ref{phi6} and  \ref{phi10}, we confirm that the hybrid stars with the same maximal mass have much smaller radius than the $\phi^4$ case. For~example, considering the curves with the maximal mass 2.6 $M_{\odot}$, we find that the radius is 16 $\textrm{km}$ for $\phi^6$ model and 13 $\textrm{km}$ for $\phi^{10}$ one, which are all much shorter than the 25~$\textrm{km}$ of $\phi^4$ model.

In Figure~\ref{phi6}, it shows the mass--radius relations for the hybrid stars of scenario III made of nuclear matter of SLy4 EoS and dark matter of $\phi^6$ model's EoS with ${\cal B}_6=0.011$ (brown lines) and $0.008$ (green lines). The~pure SLy4 neutron stars are marked in red as before. Similarly, the~neutron and dark branches are denoted by solid and dash-dotted lines, respectively; and then we apply the (reverse) BTM criteria to judge the saddle (in-)stabilities as before. We find that the $BDA$ line marked in Figure~\ref{phi6} bends higher than that in Figure~\ref{phi4}, but~still point $A$ is unstable because of the existence of the minimum point $D$, as~Section~\ref{sec 3} tells us. Again, the~stable regions are near the pure star configurations, and~for the case of ${\cal B}_6=0.011$ are circled by the blue closed path.  { Again,
 we find that the maximal mass of the stable hybrid stars of this kind cannot be higher than the maximal mass, i.e.,~2.1 $M_{\odot}$ of pure neutron stars, similar to the $\phi^4$ case. }

{ However, we see the above mass bound is lifted when considering the hybrid stars of scenario III made of nuclear matter SLy4 EoS and the dark matter of $\phi^{10}$ model's EoS, and~the maximal mass can be around 2.6 $M_{\odot}$ to explain GW190814. We have also considered the hybrid star configurations of scenario III for $\phi^8$ model (not shown here) and reach a maximal mass about 2.3 $M_{\odot}$.} The mass--radius relations are shown in Figure~\ref{phi10} for $\phi^{10}$ model's EoS with ${\cal B}_{10}=0.0036$ (brown lines) and $0.0028$ (green lines). The~line styles are the same as in Figures~\ref{phi4} and  \ref{phi6}. The~main reason for the lifting of the mass bound is the disappearance of the local minimum $D$ in Figures~\ref{phi4} and  \ref{phi6}, and~now we see that the intersection point $A$ in Figure~\ref{phi10} is no longer a saddle point so that the stable region is extended beyond the nearby region of the pure neutron star curve (the red curve). The~stable region for ${\cal B}_{10}=0.0036$ is again circled by the blue closed path, however, in~which we see the maximal mass is about  2.6 $M_{\odot}$. This can then be used to explain the heavy companion of the black hole in~GW190814.

{Bosonic $\phi^4+\phi^6$ model. } 
The EoS for this model is given in \eq{EoS46-1} and \eq{EoS46-2}. Naively, it seems to be the intermediate model between $\phi^4$ and $\phi^6$ models. However, as we have seen in Section \ref{sec 2}, it contains two tuning parameters $\cal B$ and $\beta $ which lead to some novelty: there are some regions with $p<0$ or even $\rho<0$. We will only consider the region for positive $p$ and $\rho$.

\begin{figure}[htbp] 
\centering
\includegraphics[height=0.5\textwidth]{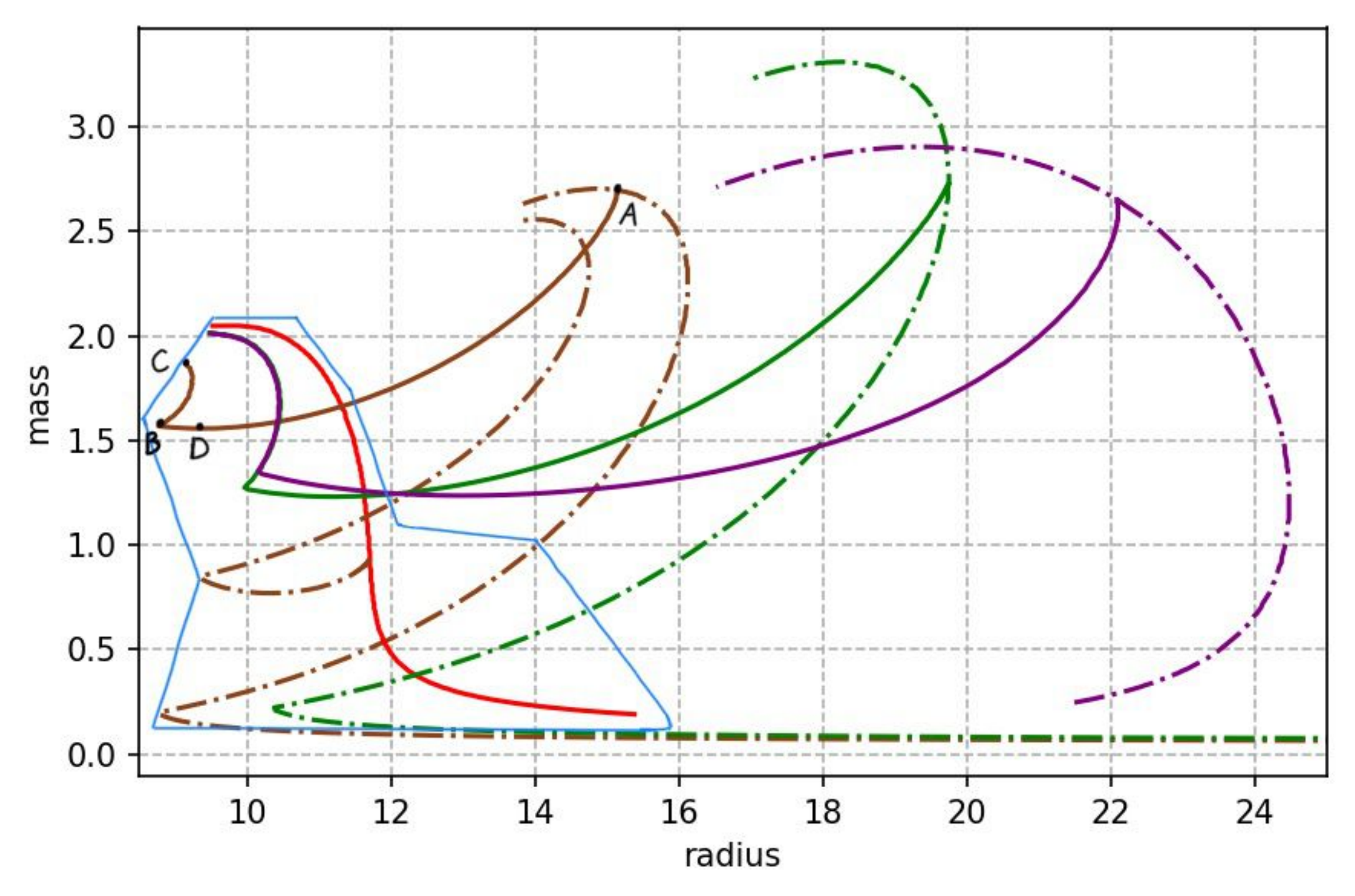}
\caption{Mass--radius relation for the hybrid stars of scenario III, with SLy4 EoS for neutron and $\phi^4+\phi^6$ model EoS given by \eq{EoS46-1} and \eq{EoS46-2} for dark matter. The line styles are the same as in Figure~\ref{phi4} but with the brown lines for ${\cal B}=0.0006$, ${\beta}=0.1$, the green lines for ${\cal B}=0.0004$, ${\beta}=0.1$, and purple lines for ${\cal B}=0.0006$, ${\beta}=-0.1$. The blue-circled regions roughly indicate the stable hybrid star configurations for ${\cal B}=0.0006$, ${\beta}=0.1$. Saddle instabilities can be checked at each intersection point of solid and dash-dotted lines with the same color. The unit of mass is $M_{\odot}$, and the unit of radius is km.}
\label{phisix}
\end{figure}

In Figure \ref{phisix}, it shows the mass--radius relations for the hybrid stars of scenario III made of nuclear matter of SLy4 EoS and dark matter of this EoS with ${\cal B}=0.0006$, ${\beta}=0.1$ (brown lines), ${\cal B}=0.0004$, ${\beta}=0.1$ (green lines) and ${\cal B}=0.0006$, ${\beta}=-0.1$ (purple lines). The line styles are the same as in Figure~\ref{phi4}. We then run through the same check of saddle (in-)stabilities as before to pin down the stable and unstable regions by applying the (reverse) BTM criteria. As a result, we find that the maximal mass is mainly affected by the value of ${\cal B}$, with lower ${\cal B}$ leads to higher mass, similar to the behavior of ${\cal B}_n$. On the other hand, $\beta$ affects the compactness, namely, the radius becomes smaller as $\beta$ grows from negative values to the positive ones. However, the change of radius  significantly slows down when $\beta \geq 0.1$.  Though the $\beta=0$ case is not included in Figure \ref{phisix}, this case is equivalent to $\phi^6$ model with ${\cal B}_6=(3 {\cal B}^2)^{1/3}$ as discussed in Section \ref{sec 2}. For example, ${\cal B}=0.0006$ corresponds to ${\cal B}_6=0.010$, which is comparable to the ${\cal B}_6=0.011$ case (brown lines) in Figure \ref{phi6} .

As shown in Figure \ref{phisix}, although the EoS becomes considerably stiffer when $\beta$ increases, we still cannot have stable stars around 2.6 $M_{\odot}$ because there is a local minimum at $D$ so that $A$ is a saddle point. The blue closed path roughly encloses the stable region for ${\cal B}=0.0006$, ${\beta}=0.1$. By varying $\cal B$ and $\beta$, the maximal mass of the stable region is about 2.4 $M_{\odot}$, when ${\cal B}=0.0007$ and ${\beta}=0.1$.

\section{Conclusions} \label{sec 5}
In this paper, we have extended the usual study of dark and hybrid stars for $\phi^4$ SIDM to more general types of bosonic SIDM models by extracting their EoSs in the isotropic limit so that we can have more access to the complete mass--radius relations for the TOV configurations. Among them, we are especially interested in the $\phi^n$ models which can have a stiff EoS when $n$ is large enough, say~$n=10$. These kinds of models can be motivated by the UV $Z_n$ flavor symmetry, which may be a natural symmetry for some higher theories. It is fascinating to further explore this connection between particle physics and the resultant astrophysical compact objects in the future. 

In general, it is easy to tune the parameters in SIDM to yield compact objects with masses comparable to or higher than the ones of neutron stars. Therefore, it is interesting to check the dark star possibilities by the future observations of the compact objects via the gravitational wave observations. Similar conclusions can be reached for the hybrid stars of scenarios I and II made of nuclear and bosonic SIDMs. In particular, many such kinds of dark and hybrid stars can have masses more than 2.6 $M_{\odot}$, and thus can be adopted to explain the recent GW190814 in which one companion compact object with such a mass has been identified by parameter estimation, which is hard to be explained by the usual EoSs for neutron stars.

On the other hand, the hybrid stars of scenario III are subjected to the saddle instability, and it is difficult for such kind of stars to have higher mass such as 2.6 $M_{\odot}$. However, in this paper, we do find such a stable massive configuration when the $n$ of the $\phi^n$ model rises up to $10$ for which the saddle instability around 2.6 $M_{\odot}$ is lifted. Although in practice it is hard to tell if a compact object can be the hybrid stars of scenario III in the near future, we may hope that the obstacle will be overcome in the long run to have precise measurements on the structure of the stars via gravitational wave detection to pin down the scenarios. Despite that, theoretically, it is interesting to see the existence of some mass bound by the saddle instability. In this paper, we simply assume that the BTM criteria still work for the mixing fluids without the mutual interaction, applying it to judge the saddle instability. It will be illuminating to study the saddle instability by directly examining the eigenspectrum of the radial oscillation modes. 

Overall, our work demonstrates the intriguing interplay between the particle physics models of dark matter and the astrophysical observations via the gravitational wave detection.  We hope this will encourage more works to explore the dark matter physics via the study of dark and hybrid stars in the new era of gravitational astronomy.

~
\section*{Acknowledgement}
FLL is supported by Taiwan Ministry of Science and Technology (MoST) through Grant No.~109-2112-M-003-007-MY3. KZ (Hong Zhang) is supported by MoST through Grant No.~107-2811-M-003-511. We thank Guo-Zhang Huang and other TGWG members for helpful discussions. We also thank NCTS for partial financial support.







\end{document}